# Improving $Nb_3Sn$ Cavity Performance Using Centrifugal Barrel Polishing

Eric Viklund
*Department of Materials Science and Engineering, Northwestern University and Fermi National Accelerator Laboratory*

David N. Seidman
*Department of Materials Science and Engineering, Northwestern University*

David Burk and Sam Posen
*Fermi National Accelerator Laboratory*

In this study we will show a new method of polishing for $Nb_3Sn$ cavities known as centrifugal barrel polishing (CBP). Using this method, $Nb_3Sn$ coated samples are polished to a surface roughness comparable to a traditional Nb cavity after electropolishing (EP). We also investigate different methods of cleaning the $Nb_3Sn$ surface after CBP to remove residual abrasive particles. The polished $Nb_3Sn$ surface is analyzed using confocal laser microscopy, and scanning electron microscopy (SEM) is used to image the surface and measure the surface roughness after polishing. Transmission electron microscopy (TEM) is also used for high resolution analysis of the surface after polishing. Finally, we show that centrifugal barrel polishing can improve the performance of a $Nb_3Sn$ SRF cavity.

## I. INTRODUCTION

Superconducting radiofrequency (SRF) cavities are an essential component of particle accelerators used in various fields of science and industry, including high-energy physics, material science, and medical applications. These cavities accelerate charged particles to very high energies and can achieve higher accelerating gradients than normal conducting cavities.

The performance of superconducting radio-frequency cavities is determined by the superconducting properties of the surface layer of the cavity. The low-temperature superconductor $Nb_3Sn$ has a higher superconducting transition temperature ($T_c$), 18 K compared to 9 K, and a higher super-heating magnetic field ($H_{sh}$), around 440 mT compared to 250 mT for the more commonly used niobium[1]. SRF cavities coated with a layer of $Nb_3Sn$ can, therefore, achieve a much higher accelerating field, up to 100 MV/m in theory, and have a lower surface resistance than niobium SRF cavities at higher temperatures allowing for cavity operation at 4 K instead of the standard 2 K operating temperature. These properties make $Nb_3Sn$ SRF cavities a promising research topic for future accelerators such as high-energy linacs or small-scale industrial accelerators.

$Nb_3Sn$ cavities are typically manufactured by coating a Nb cavity with a thin film of $Nb_3Sn$ using Sn vapor-diffusion, exposing a niobium cavity to tin vapor at 1,100 °C to create $Nb_3Sn$. The reaction forms a 2-3 µm thick $Nb_3Sn$ film with grains approximately 1 µm large. The grains are faceted, and the grain boundaries are thermally etched, resulting in approximately 100-150 nm of surface roughness.

Using the Sn vapor-diffusion coating technique, $Nb_3Sn$ cavities have only been able to reach a maximum accelerating gradient of 24 MV/m[2], a result that was achieved using a thinner $Nb_3Sn$ coating which is smoother than the typical coating. However, this result has not been reproducible and the performance is lower than the theoretical maximum of $Nb_3Sn$ cavities.

Surface roughness is thought to be one of the limiting factors of $Nb_3Sn$ SRF cavity performance. Simulations of the magnetic field near a typical $Nb_3Sn$ surface show that the magnetic field is increased by up to 60 percent in some areas compared to a smooth surface[3] and could be higher for particularly rough areas. By reducing the field enhancement caused by surface roughness, a corresponding increase in the cavity accelerating gradient can be expected.

One method to achieve smoother $Nb_3Sn$ surfaces is polishing. $Nb_3Sn$ polishing has been a topic of study for some time. Chemical methods such as electropolishing (EP)[4–6], buffered chemical polishing (BCP)[5, 6], and oxy-polishing[4, 5] have been studied, but have failed to produce any meaningful improvement in surface roughness. The expected reason for this is that a large amount of material removal is required to produce a substantial smoothing effect when utilizing chemical methods. Electropolishing treatments for Nb typically remove between 5 µm and 10 µm of material to achieve a smooth surface depending on its initial roughness. This amount of material removal is infeasible for $Nb_3Sn$ films, since their thickness is only 2-3 µm. Additionally, niobium and Sn react differently to the chemicals used, which can lead to different removal rates for each element, thus changing the surface stoichiometry. Even a small change in the stoichiometry away from $Nb_3Sn$ can cause a decrease in $T_c$[7].

Centrifugal barrel polishing (CBP) is another method used to polish SRF cavities, which utilizes an abrasive material to mechanically smooth the surface. This method has been used to repair surface damage and to attain a very smooth surface in Nb SRF cavities[8]. As of yet, only very limited attempts to apply this technique to $Nb_3Sn$ cavities have been made.

In this paper, we show a procedure for mechanically polishing $Nb_3Sn$ cavities using CBP. First, $Nb_3Sn$ coated samples are polished to determine the effectiveness of the CBP method and to determine the optimum polishing parameters such as tumbling duration and the abrasive material. The results of the sample experiments are used to decide the polishing parameters for a $Nb_3Sn$ coated, 1.3 GHz, TESLA geometry SRF cavity. The RF performance of the cavity is tested before and after the CBP treatment. Finally, the cavity is treated with a low temperature Sn coating process to repair the surface and the RF performance is once again tested.

## II. SAMPLE STUDY

Since $Nb_3Sn$ is a relatively unexplored material, there are no established polishing parameters or abrasive materials to achieve a good surface finish. To allow for rapid iteration and microscopy surface analysis, we first perform polishing experiments on $Nb_3Sn$ samples. To evaluate the performance of CBP, the surface roughness of the polished samples is measured using confocal laser microscopy and the surface is analyzed using scanning and transmission electron microscopy (SEM and TEM). The material removal rate is measured using focused ion-beam tomography.

### A. Centrifugal Barrel Polishing

Centrifugal Barrel Polishing (CBP) is a technique that was developed by Cooper and Cooley[8] as a method of polishing Nb cavities without using toxic chemicals such as HF. The technique uses a custom built tumbling machine that can fit up to 9-cell size 1.3 GHz cavities. When a cavity is mounted in the tumbling machine and filled with abrasive slurry, the rotating motion of the cavity accelerates the polishing media against the cavity surface with up to 6 g of force.

The abrasive material determines the removal rate and minimum surface roughness attainable using CBP. Large-grit material is used to remove material quickly and smooth out large defects like pits and scratches while fine-grit material is used to microscopically smooth the surface. The removal rate of different abrasive materials has been studied by Palczewski, et. al.[9]

Since the roughness of as-coated $Nb_3Sn$ cavities is on the order of 100-200 nm, our experiments focus on using fine-grit materials. In this experiment, we use a colloidal nano-particle suspension as our abrasive material. 50 nm diameter alumina and 40 nm diameter silica nanoparticles suspended in water were tested, but we found no discernible difference between the two materials.

The nano-particle suspension was mixed with a large, soft material to act as a carrier. The purpose of the carrier material is to carry the nano-particles and to apply a force between them and the cavity surface. Two carrier materials were tested, 13 mm diameter wooden balls and 25 mm felt cubes.

### B. Coupon Cavity

To test the centrifugal barrel polishing method on $Nb_3Sn$ samples in a realistic environment, we use a coupon cavity. This cavity has multiple ports where samples are mounted. The samples sit flush with the inside surface of the coupon cavity, as is shown in Fig. 1, where they experience identical polishing conditions to a real cavity surface. This allows for sample experiments that are representative of the final cavity polishing process. Using this method we inspect the $Nb_3Sn$ surface after polishing under a microscope to determine the best polishing parameters.

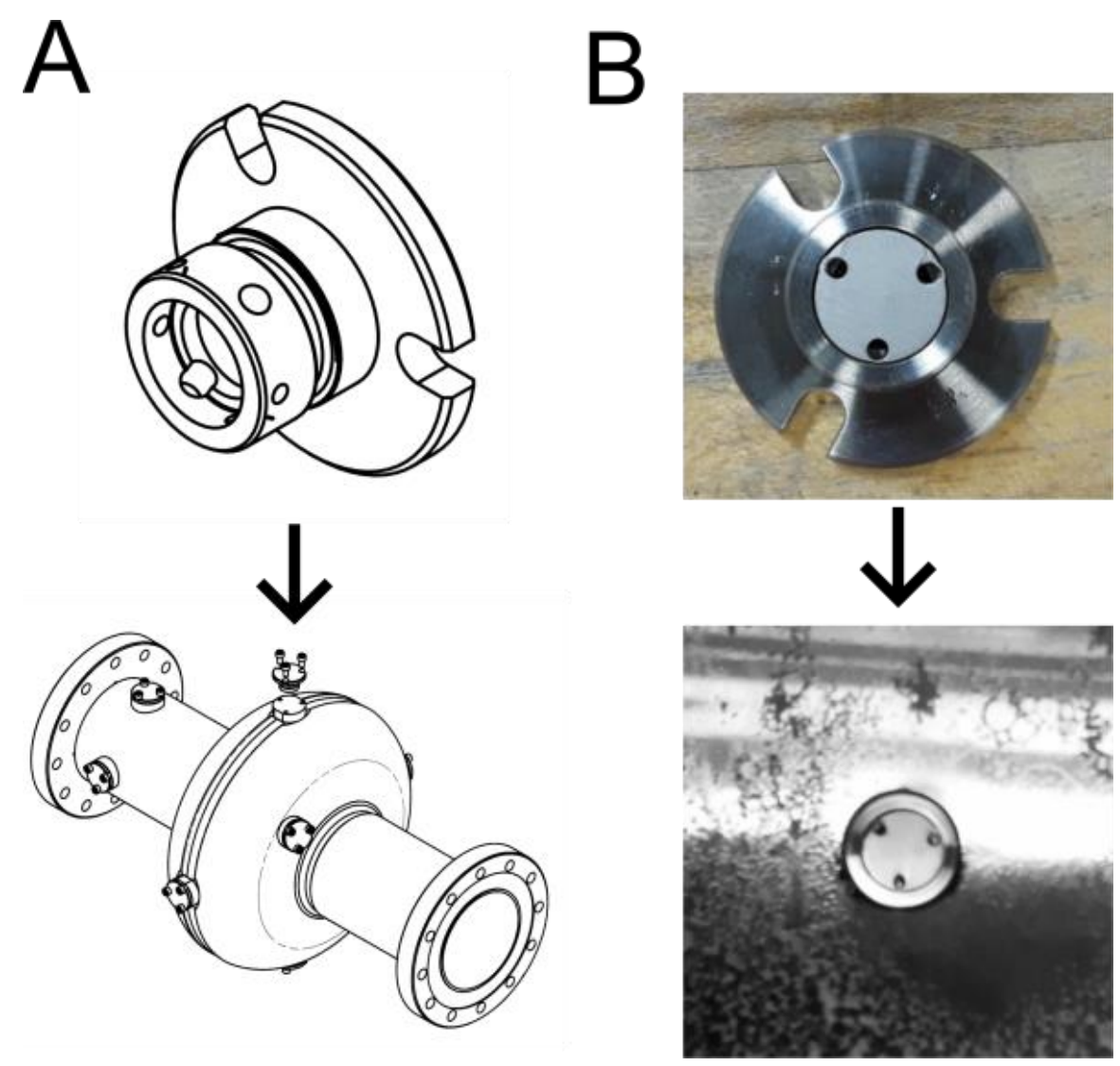

FIG. 1. (A) A schematic of the coupon cavity and the sample holder used to polish the $Nb_3Sn$ coated samples. The sample holder can hold 1 cm diameter disks by clamping the sides of the sample with set screws. (B) Pictures of the sample holder sitting outside the coupon cavity with a sample mounted and as seen from the inside of the coupon cavity.

### C. $Nb_3Sn$ Coating Using Sn Vapor-Diffusion

The $Nb_3Sn$ samples and $Nb_3Sn$ cavity used in this study were coated at Fermilab in a high-vacuum furnace. The coatings were created at 1,100 °C with a Sn crucible acting as the Sn source as well as $SnCl_2$ acting as a nucleating agent. A detailed review of the coating system at Fermilab shows the specific operating details of the coating system[10].

### D. Surface Analysis of Mechanically Polished $Nb_3Sn$ Coated Samples

The $Nb_3Sn$ samples were polished for different lengths of time ranging from 2 to 8 hours using the wooden spheres or the felt cubes as the carrier material. A height map of the polished samples is shown in Fig. 2. The smoothness of the samples clearly improves as longer polishing is applied.

By comparing the surface optical micrographs over time, it is clear that material is preferentially removed from the highest point on the surface, causing the sharp peaks on the surface to be removed quickly while valleys in the surface are left untouched. This is different from EP, which preferentially smooths areas with high curvature including both peaks and valleys. Due to this different smoothing mechanism, surface roughness is minimized when the thickness of material removed is equal to the height difference between the highest and lowest point on the surface, which is around 1 μm. This is confirmed by the sample experiments, after 8 hours of polishing only the deepest valleys of the initial coating remain.

The surface height maps are used to calculate the root mean square (RMS) surface roughness of the samples and is shown in Fig. 3. After 6 hours of polishing the surface roughness is comparable to the surface roughness of the well performing, thinly coated $Nb_3Sn$ coatings created at FNAL[2]. After 8 hours of polishing, the surface roughness is comparable to a typical Nb surface after EP. This level of smoothness has never been achieved for $Nb_3Sn$ cavities until now. At this level of surface roughness, the performance degradation caused by field enhancement due to surface roughness should be greatly reduced.

The thickness of the film is measured after polishing using FIB/SEM. Our measurements show that only a small amount of material is removed even after 8 hours of polishing. Samples polished using the felt media show an average removal rate of 170 nm/hour whereas the wooden media shows an average of 95 nm/hour removal rate. The starting thickness of the samples is between 3-3.5 μm. After 8 hours of polishing there is still over 1.5 μm of $Nb_3Sn$ left on the surface. To completely shield the Nb substrate from the RF fields only a few hundred nanometers of material are required, since the London penetration depth of $Nb_3Sn$ is approximately 100 nm[1].

The surface of the polished samples was analyzed using SEM and TEM to look for surface damage or chemical changes on the surface caused by the tumbling or cleaning process. As seen in Fig. 6 and Fig. 7, the $Nb_3Sn$ samples polished using wooden spheres were damaged resulting in microscopic scratches on the surface and a damaged layer consisting of a nanometer-scale disordered $Nb_3Sn$ layer. No surface damage was detected on samples polished using the felt cubes. Since the surface damage may negatively affect the cavity performance, the felt cubes are best to use for polishing $Nb_3Sn$.

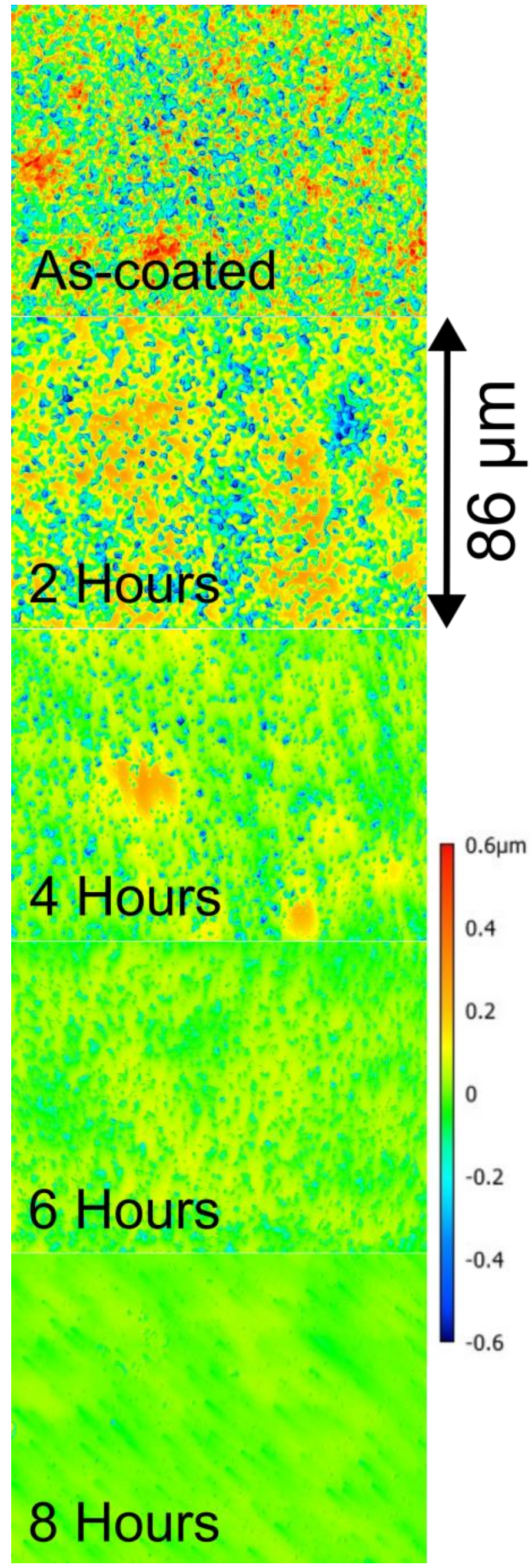

FIG. 2. Surface height maps of $Nb_3Sn$ samples mechanically polished for different lengths of time ranging from 2 to 8 hours compared to the initial state of the $Nb_3Sn$ coating.

## III. POLISHING A $NB_3SN$ CAVITY USING CBP

Given that CBP was able to produce a smooth surface on $Nb_3Sn$ samples, the next step is to apply the polishing to a $Nb_3Sn$ cavity. A $Nb_3Sn$-coated cavity was polished using the felt cube polishing media with a 50 nm alumina abrasive particle suspension, chosen to avoid the risk of Si contamination in the coating furnace. The cav-ity was polished for 4 hours followed by high-pressure water rinsing and ultrasonic cleaning for 30 minutes to remove any residual abrasive material left by the polishing process. These parameters were chosen as a conservative estimate to minimize the possibility of removing the $Nb_3Sn$ film and allow for more material removal in the future while still providing a considerable improvement in surface roughness.

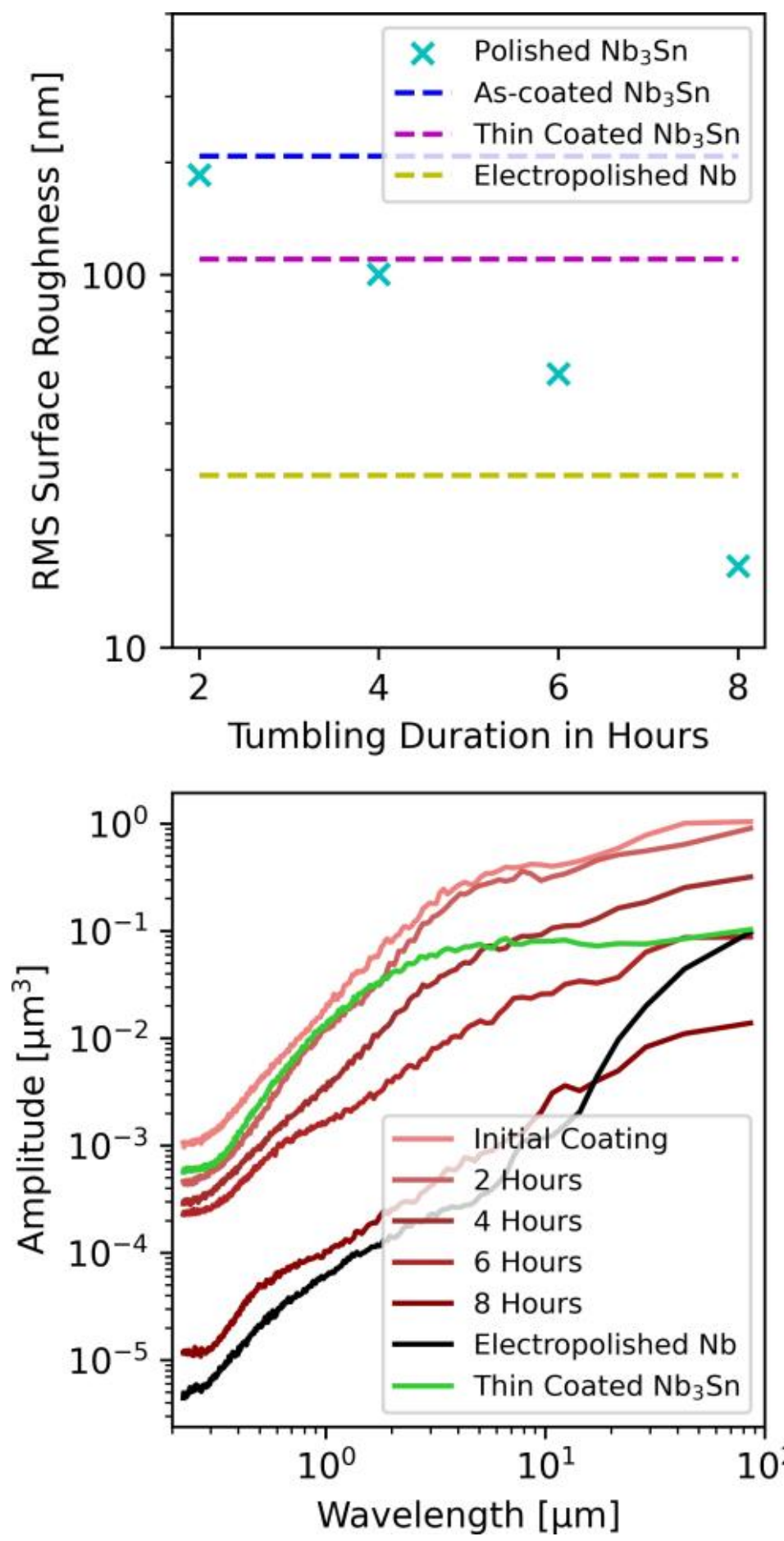

FIG. 3. Surface roughness of $Nb_3Sn$ samples mechanically polished for different lengths of time calculated from the surface height maps (top). The power spectral density (PSD) of the surface profile after different amounts of tumbling as well as the PSD of electropolished Nb and a thinly coated $Nb_3Sn$ film (bottom).

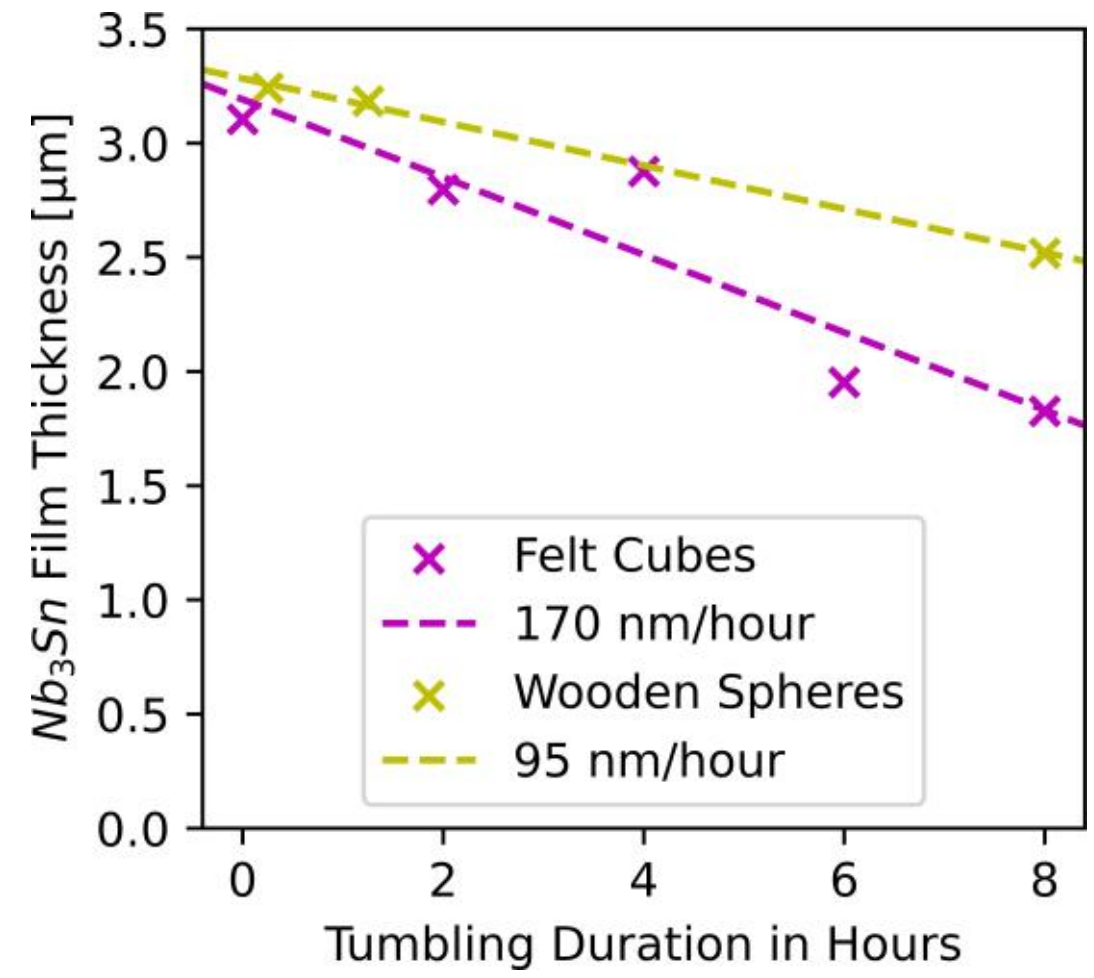

FIG. 4. The thickness of the $Nb_3Sn$ film after mechanically polished for different lengths of time using felt cubes or wooden spheres as the polishing media.

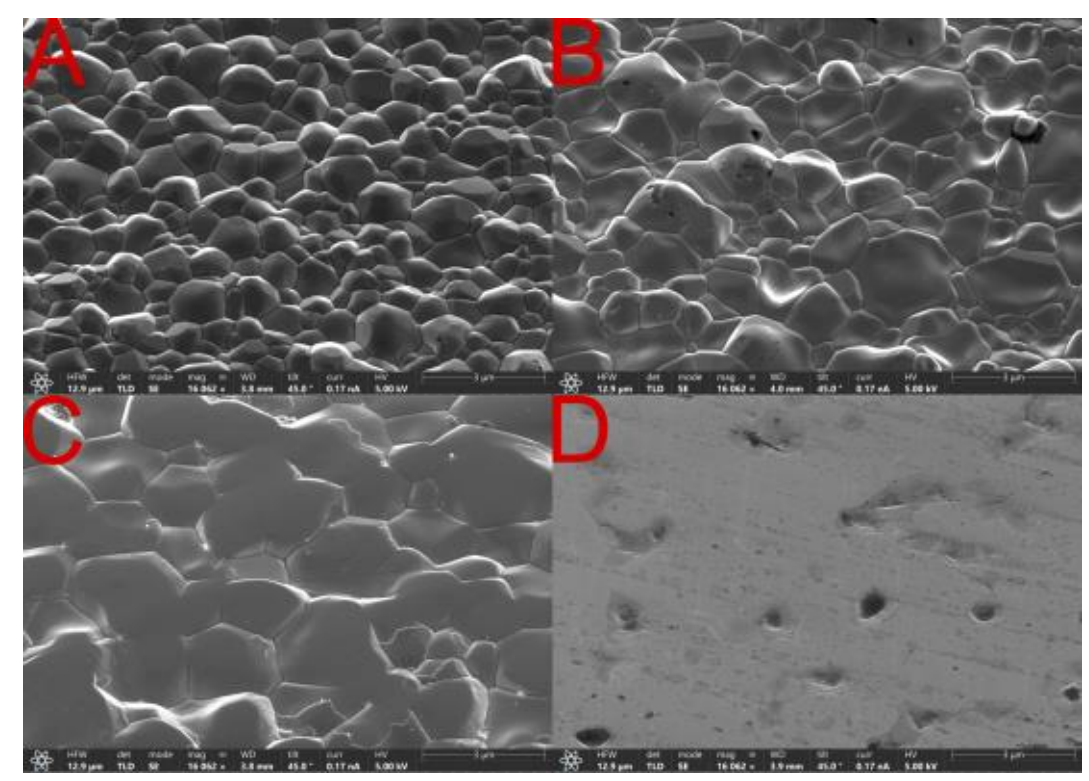

FIG. 5. SEM micrographs of a $Nb_3Sn$ a thin coated sample (A), standard coated sample (B), a sample after polishing for 2 hours (C), and a sample after polishing for 6 hours (D).

Visual inspection of the cavity shows that the surface roughness was improved by the polishing procedure. The as-coated surface of the cavity has a matte finish, which is common on $Nb_3Sn$-coated surfaces, and after the polishing the cavity has a shiny surface finish. This is indicative of the removal of microscopic surface roughness on the surface of the cavity.

### A. Low Temperature Recoating Procedure

After the $Nb_3Sn$-coated cavity was polished using CBP, a secondary coating was applied, which we refer to as the recoating procedure. The purpose of this coating is to repair any surface damage caused by CBP or any subsurface defects, such as tin-deficient regions, that may have been exposed.

The recoating procedure was performed at 1,000 °C. This lower temperature was chosen to minimize any thermal etching of the surface, which could increase the surface roughness. No $SnCl_2$ was used as it is unnecessary to nucleate any $Nb_3Sn$ grains. One third of the normal amount of Sn was used during the coating, since no additional film growth is needed. The coating was performed at Fermilab using the coating furnace mentioned in Section II C.

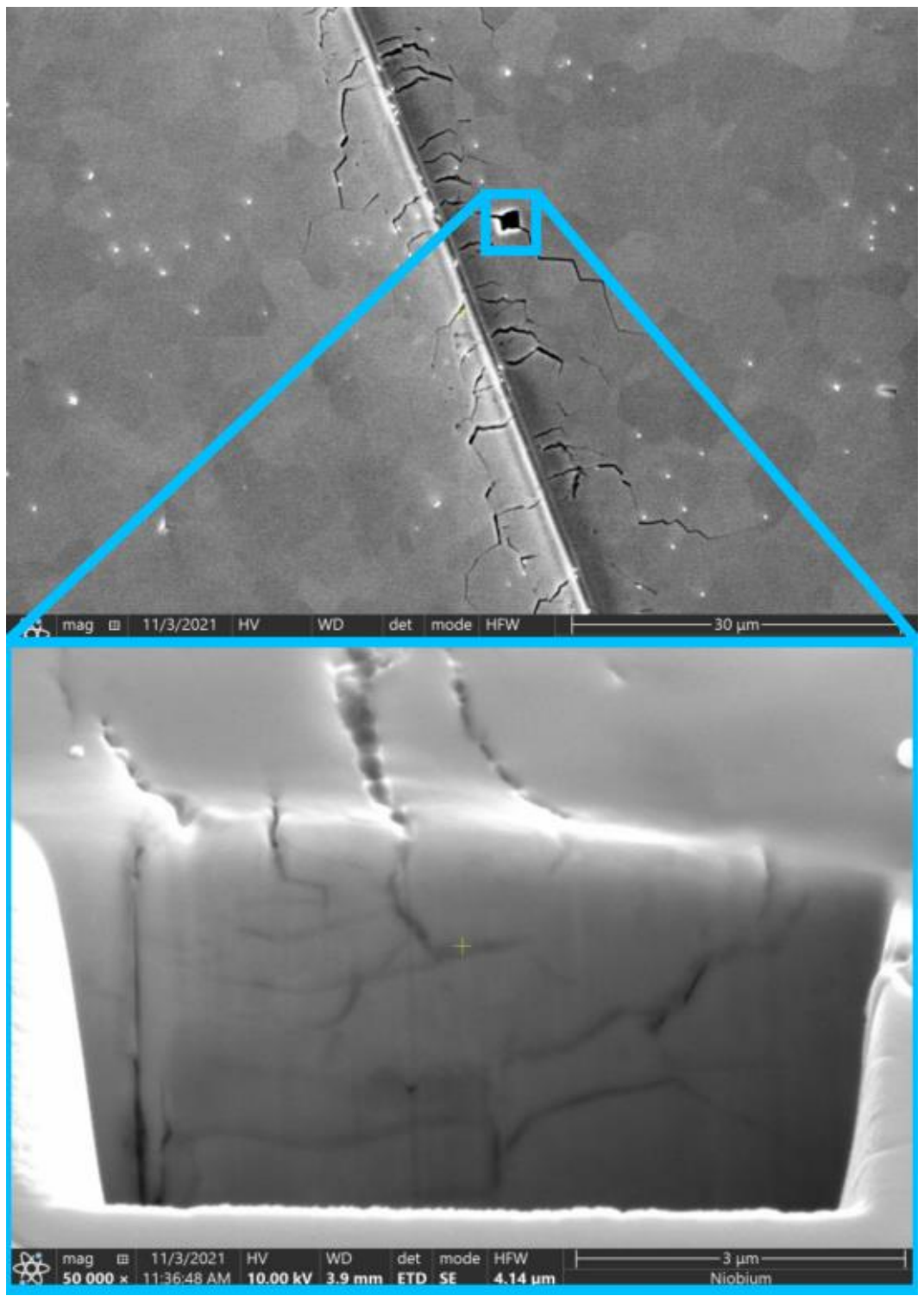

FIG. 6. SEM micrograph showing a $Nb_3Sn$ sample polished for 30 hours using wooden spheres and a colloidal abrasive suspension. $Nb_3Sn$ films polished using wooden spheres show microscopic scratches and cracks on the surface. A square hole is cut into the surface, visible in the top micrograph to expose a cross section of a crack. The cross section shows that the cracks penetrate deep into the film.

### B. SRF Cavity RF-Performance Testing

The RF performance of the $Nb_3Sn$-coated cavity was tested three times; first, in the as-coated state with no polishing applied; second, after the CBP treatment; lastly, after the recoating procedure. The performance was tested using the vertical test stand (VTS) at FNAL[11].

### C. Testing the Polished $Nb_3Sn$ SRF Cavity

The as-coated performance of the cavity was poor compared to most other $Nb_3Sn$ cavities reaching an accelerating gradient of around 10 MV/m with a Q of $10^{10}$ at 4.4 K. After the polishing is applied, the cavity exhibits Q-slope (the quality factor decreases with increasing accelerating field), and the maximum gradient was only 5 MV/m.

After the cavity was treated with the recoating treatment, detailed in SectionIII A, the Q-slope is ameliorated and the maximum accelerating gradient increases to 15 MV/m. The quality factor of the cavity was also improved over the as-coated state at 2.0 K, but not at 4.4 K. Fig. 8 shows the performance of the cavity after each step.

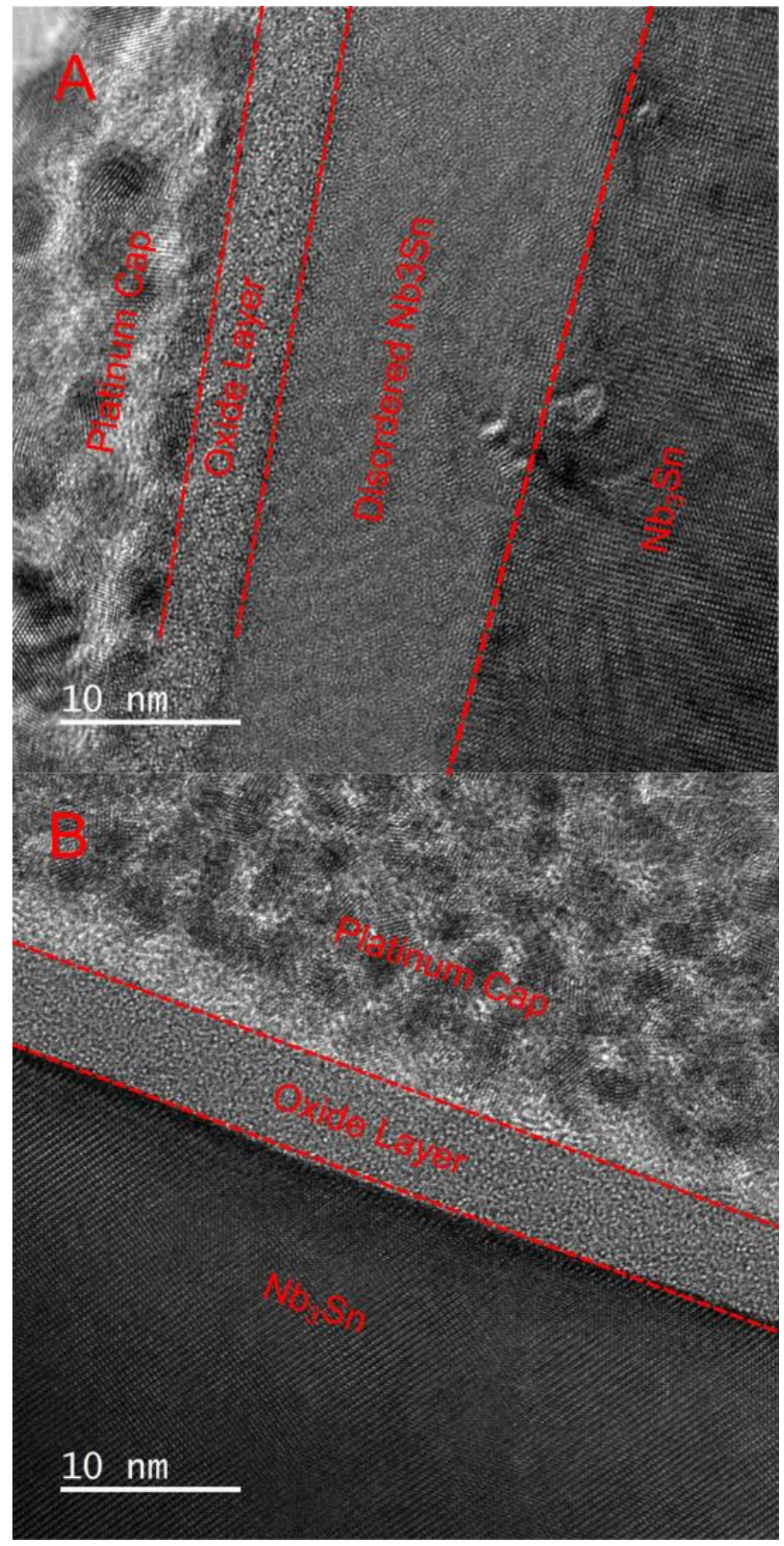

FIG. 7. TEM images of a $Nb_3Sn$ sample polished using wooden spheres (A) and felt cubes (B). The polishing procedure creates a 10 nm thick layer of disordered $Nb_3Sn$ on the sample polished with wooden spheres which is not present on the sample polished by felt cubes.

## IV. DISCUSSION

Using mechanical polishing, we are able to produce smooth $Nb_3Sn$ films with surface roughness less than 20 nm. This level of surface roughness has thus far been impossible to achieve using existing methods.[2, 4–6] We have also shown that this smoothing can be achieved with only a few hundred nanometers of material removal, much less than what would be required for chemical polishing methods.

This study also shows that mechanical polishing can be used to improve the performance of $Nb_3Sn$ cavities when used in conjunction with a recoating procedure. The immediate effect of polishing reduces the quality factor and accelerating gradient of the cavity. The cause of this performance degradation is not known, but could be due to subsurface defects exposed to the surface such as tin-depleted regions[12]. In a previous study, we have shown that tin-depleted regions and regions where the film becomes very thin are common in tin vapor-diffusion coated samples[13].

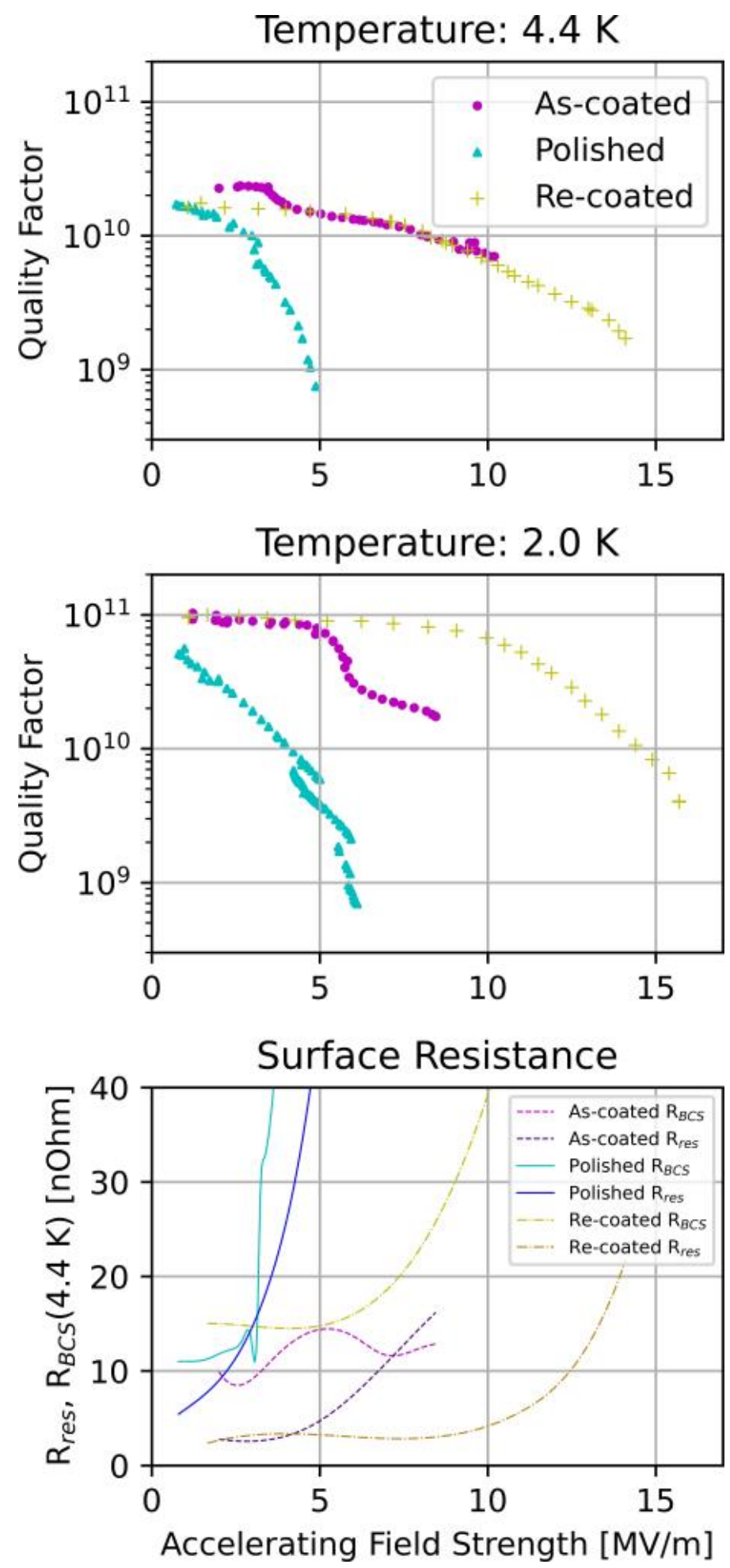

FIG. 8. The RF performance of the $Nb_3Sn$-coated SRF cavity before and after mechanically polished and after a recoating treatment. The residual resistance of the cavity is calculated from the 2.0 K measurements under the assumption that the BCS resistance is negligible at this temperature. We assume that the additional resistance measured at 4.4 K is entirely BCS resistance. These assumptions may be false if there are multiple $Nb_3Sn$ phases on the surface.

Another possible cause for performance degradation is surface damage or contamination caused by the polishing process. Residual abrasive particles can be seen in the SEM images of the polished samples. Cleaning the surface removes most of the contamination, but there may still be some abrasive particles left on the surface, which could cause performance degradation. Surface damage such as cracks or scratches on the film could also degrade performance, although this seems unlikely as no surface damage was detected on the $Nb_3Sn$ samples polished with felt cubes. It is possible that the oxide that forms on the Nb3Sn surface after polishing is unfavorable for performance compared to the oxide that forms in the furnace after the coating process. More work is required to determine the cause of the performance degradation. After the recoating process was applied, the cavity performance was improved over the unpolished state. We theorize that the re-coating procedure eliminates the performance degradation by repairing any defects exposed to the surface by the polishing step. A short, low-temperature coating is sufficient to diffuse more Sn into any exposed tin-depleted regions, repair small cracks, and return the surface oxide of the film to its as-coated state. However, it is difficult to determine the exact effects of the recoating without performing a thorough cutout analysis on the cavity. We plan to conduct studies in the future to determine the cause of the performance degradation after mechanical polishing and the effects of the recoating procedure on the $Nb_3Sn$ surface.

Despite the improvement to surface roughness, the accelerating gradient of the mechanically polished cavity is still below that of the current record holding cavity.[2] This suggests that there are multiple mechanisms that contribute towards cavity quench with surface roughness being one of them. By using mechanical polishing to eliminate the effects of surface roughness on the cavity performance, we can isolate these other causes and study them individually. We plan to apply mechanical polishing to $Nb_3Sn$ cavities with better as-coated performance so that the differences between these cavities can be determined. Applying the polishing procedure to a well-performing cavity could potentially lead to a drastic increase in the maximum accelerating gradient of $Nb_3Sn$ cavities if the performance improvement shown in this paper can be applied to well performing as-coated cavities.

The ability to smooth the surface of the $Nb_3Sn$ cavities also opens opportunities to experiment with thicker coatings, which can then be polished for a longer duration to achieve a smoother surface. Future research will include a study on polishing thicker $Nb_3Sn$ coatings and analyzing the prevalence of defects in these films.

## V. CONCLUSION

The work presented in this paper shows that centrifugal barrel polishing is a promising treatment for $Nb_3Sn$ cavities. Through a series of sample studies, we were able to develop a mechanical polishing procedure that can produce surface roughness that was previously un-obtainable. We have found that it is possible to attain films with a surface roughness of 20 nm or lower.

Furthermore, we have shown that this surface polishing technique can be used to improve the performance of $Nb_3Sn$ coated cavities when it is paired with a recoating step, which consists of a short, low-temperature Sn coating.

## VI. ACKNOWLEDGEMENTS

This manuscript has been authored by Fermi Research Alliance, LLC under Contract No. DE-AC02-07CH11359 with the U.S. Department of Energy, Office of Science, Office of High Energy Physics.

This work made use of the EPIC facility of Northwestern University's NUANCE Center, which has received support from the SHyNE Resource (NSF ECCS-2025633), the IIN, and Northwestern's MRSEC program (NSF DMR-1720139).